\documentclass{appolb}
\usepackage{graphicx}
\usepackage[dvipsnames]{xcolor}

\begin{document}
\title{General unquenching properties \\ %
of two-meson scattering %
and production amplitudes %
\thanks{Presented at Excited QCD 2017,
Sintra (Portugal), 7--13 May 2017}%
}
\author{Eef van Beveren
\address{Centro de F\'{\i}sica da UC,
Departamento de F\'{\i}sica,\\ Universidade de Coimbra,
P-3004-516 Coimbra, Portugal}
\\ [10pt]
George Rupp
\address{Centro de F\'{\i}sica e Engenharia de Materiais Avan\c{c}ados,\\
Instituto Superior T\'{e}cnico,
Universidade de Lisboa,
P-1049-001 Lisboa, Portugal}
}
\maketitle
\begin{abstract}
Besides the unitarity and symmetry requirements
for a multi-resonance scattering amplitude,
several other natural conditions can easily exclude unrealistic proposals.
In particular, the behaviour of singularities
under the variation of model parameters yields important information.
We discuss how resonance poles should move in the complex-energy plane
when coupling constants and masses are varied,
how resonances above threshold can turn into bound states below threshold,
and how the light-quark spectrum
can be turned into the spectrum of heavy quarks,
with one and the same analytic expression for the scattering amplitude.
Moreover, it is shown that perturbative approximations
usually do not satisfy these natural conditions.
\end{abstract}

\section{Introduction}

It was a pleasure to participate in the Excited QCD 2017 workshop
at Sintra (Portugal)
and to attend short seminars on so many different approaches
towards understanding the complicated relation between
strong interactions and quantum chromodynamics (QCD).
In particular, we were pleased to witness presentations
on the study of hadron scattering, mass distributions,
cross sections, and resonances.
Nevertheless, although it did not seem to disturb most theoreticians,
the lack of progress in measuring high-statistics
multi-hadron data was very disappointing.
Experiment does not yet provide the necessary conditions
to confront model results with measured multi-hadron mass distributions.
Several decades of accelerators and sensitive detectors
have, unfortunately, still not resulted in sufficient data
to allow for narrow binning and high statistics.
From Fig.~1 in Ref.~\cite{ARXIV12041984}
one could even conclude that it will take quite a while
before ATLAS statistics \cite{PRL108p152001}
competes with 25 years older data
of the ARGUS Collaboration \cite{PLB160p331}.

During the workshop some criticism arose
on the compilation of data by the Particle Data Group Collaboration (PDG)
\cite{CNPC40p100001}.
However, it would actually be more in place to direct
such criticism towards researchers who use section headings
to fit the results of their models.
Sure, the PDG reports on each significant enhancement observed
in multi-hadron mass distributions
as well as on its weighted central mass and width,
and furthermore its quantum numbers
based on available experimental data analyses.
Subsequently it introduces a new section in its Review of Particle Physics
when suspected to be different from already reported enhancements.
However, the PDG also supplies its readers with a list of published work
related to each one of the enhancements.
Serious researchers are thus free to go through
the published mass distributions and draw their own conclusions
on the nature of a certain enhancement
or, even better, compare the full multi-hadron mass distribution
with the results of their models.

\section{Dimeson channels}

Decades ago pions or kaons were scattered
from the proton's meson cloud (see e.g.\
Refs.~\cite{NPB133p490,NPB296p493})
in order to produce pion-meson or kaon-meson cross sections,
respectively.
However, a more elegant production of dimesons
stems from $e^{-}e^{+}$ scattering
(see e.g.\ Refs.~\cite{PRD79p092001,PRL98p092001}),
since via vector dominance one is then pretty sure
about the dimeson's quantum numbers.
Consequently, the resulting $J^{PC}=1^{--}$ mass distributions
could be considered backbones of mesonic spectra
and thus should have been given the highest priority
in the past.
Nevertheless, experiments for the vector charmonium spectrum
only came up with binnings of 20~MeV \cite{PRL98p092001}
(see Fig.~1 in Ref.~\cite{ARXIV10053490})
or 25~MeV \cite{PRD79p092001}
(see Fig.~3 in Ref.~\cite{CNPC35p319}),
and extremely low statistics,
whereas light-dimeson spectra are not in a better shape
(see e.g.\ Ref.~\cite{PRD78p072006}).

Previously we presented amplitudes for multi-resonance scattering
(see Eq.~2 in Ref.~\cite{APPS9p583})
and production
(see Eq.~3 in Ref.~\cite{APPS9p583}),
which were based \cite{PRD21p772}
on general scattering theory \cite{Taylor},
applied to multi-channel dispersion
in the presence of a tower of bound states and resonances,
the so-called Resonance Spectrum Expansion (RSE)
\cite{IJTPGTNO11p179,AP323p1215,AP324p1620}.
The RSE amplitude in Ref.~\cite{APPS9p583}
is restricted to single-channel scattering.
However, a multi-channel generalisation is straightforward.
For dimeson channels, the input spectrum can be determined
by the use of a bound-state model for quark-antiquark states.
The amplitudes can be then converted into mass distributions
and cross sections.
Furthermore, one can extract
the complex scattering singularities
(poles in the total invariant mass $\sqrt{s}$)
from the amplitudes and study their behaviour
under variation of the model's parameters.

Quantum states of the RSE respect total angular momentum $J$,
parity $P$, total flavour and isospin,
moreover, when applicable, also charge conjugation $C$.
However, they
do not have well-defined orbital quantum numbers,
relative angular momentum $\ell$, and radial excitation $n$,
but are rather mixtures of all possible orbital quantum numbers.
In particular, vector $S$ and $D$ states mix.
The latter phenomenon has interesting consequences,
as dominantly $D$-wave resonances are found near the input spectrum
and with small widths (a few MeV),
whereas dominantly $S$-wave states have central resonance masses
some 150~MeV or more below the masses of the input spectrum
and with relatively large widths (tens of MeV's).

Coupled channels do not exhibit enhancements at the same place.
A pole is determined by the full scattering matrix,
but enhancements also depend on the kinematics of a specific channel.
This can be nicely observed from Figs.~3 and 4 in Ref.~\cite{CNPC35p319},
where mass distributions are depicted for
$D^{\ast}\bar{D}^{\ast}$ \cite{PRD79p092001}
and
$\Lambda^{+}_{c}\Lambda^{-}_{c}$ \cite{PRL101p172001},
respectively.
Each of the two figures shows
the $\Lambda^{+}_{c}\Lambda^{-}_{c}$ threshold enhancement
and the $5S$, $4D$ charmonium resonances,
but masses are different.
Part of the discrepancy may stem from incompatibilities
between the mass normalisations of the BABAR and BELLE Collaborations,
but the larger part is due to differences in kinematics.
Moreover, in some channels no enhancement appears at all near the pole.

In Ref.~\cite{AIPCP814p143}
we compared to experiment \cite{NPB133p490,NPB296p493}
our predicted cross sections for $S$-wave isodoublet dispersion
of $K\pi$ (see Fig.~2),
$K\eta$ (see Fig.~6) and $K\eta '$ (see Fig.~7).
For $K\pi$ we showed results
for three different values of the overall coupling constant $\lambda$.
For very small values of $\lambda$,
one observes the scalar $n\bar{s}$ input spectrum.
When $\lambda$ takes about half its model value, one
notices some more structure for low invariant masses.
At the model's standard value of $\lambda$,
this structure is dominant and well in agreement
with the experimental data \cite{NPB133p490,NPB296p493}.
The behaviour of the poles under variation of $\lambda$
for the two lowest-lying $K\pi$ resonances
was also studied in Ref.~\cite{AIPCP814p143} (see Fig.~3).
The scattering pole for $K^{\ast}_{0}(1430)$
can directly be connected to the $n\bar{s}$ input spectrum.
However, the scattering pole for $K^{\ast}_{0}(800)$
does not stem from the input spectrum,
but is dynamically generated \cite{ZPC30p615}.

Under variation of $\lambda$, poles can
also move below the lowest threshold,
thus representing bound states.
The passage through threshold
is different for $S$ waves and higher waves.
This issue was studied in Ref.~\cite{LNP211p331}
(see Figs.~4.1 and 4.2).
The resonance pole for $P$- or higher-wave dispersion
moves smoothly towards threshold under variation of $\lambda$
(see Fig.~6 in Ref.~\cite{AIPCP688p88}).
Below threshold it behaves as expected for a bound-state pole.
In contrast, a complex $S$-wave resonance pole
can have a real part smaller
than the threshold value of $\sqrt{s}$,
and only end up on the real axis well below threshold,
thus representing a virtual bound state.
Thereafter, under variation of $\lambda$,
the pole moves back towards threshold
along the real $\sqrt{s}$ axis,
and only after touching threshold it turns into a true bound state
(see Fig.~5 in Ref.~\cite{AIPCP688p88},
or Fig.~1 in Ref.~\cite{EPJA31p468}).
Moreover, one can also continuously
vary quark masses and the corresponding threshold values.
Resonance poles then move smoothly
from one established resonance to another
(see Fig.~1 in Ref.~\cite{PRD74p037501}).

Any model that claims to describe resonances
of multi-hadron scattering or production
should exhibit the above properties for the corresponding resonance poles.
Such poles for perturbative scattering amplitudes
usually do not satisfy this behaviour at threshold,
as was studied in Ref.~\cite{PTP125p581} (see Fig.~5).

\section{Resonances}

In the harmonic-oscillator aproximation of the RSE (HORSE),
one can predict mass distributions for dimeson channels.
It was observed that the harmonic-oscillator frequency
can be taken the same (0.19 GeV) for all flavours.

The results for $K\pi$ \cite{ZPC30p615},
$K\eta$, and $K\eta^{\prime}$ \cite{AIPCP814p143}
were already discussed above.
In Table~2 in Ref.~\cite{AIPCP814p143}
we showed the five lowest-lying resonance poles
that we found in the scattering matrix for isodoublet $S$-wave channels.
We thus expect to find 10 plus 5 corresponding poles
in the isosinglet and isotriplet $S$-wave channels, respectively,
many more than observed in experiment \cite{CNPC40p100001}.

For vector states we also found many resonances
that have not yet been confirmed in experiment.
Our assignments for $D^{\ast}_{s}$ resonances are collected in
Fig.~1 and Table~3 in Ref.~\cite{ARXIV10112360}.
The 20-MeV binning of the data \cite{PRD80p092003}
does not allow for firm conclusions.
Higher dominantly $S$- and dominantly $D$-wave charmonium resonances
from HORSE have been reported by us at various occasions,
as e.g.\ in Fig.~5 in Ref.~\cite{CNPC35p319},
where one may observe that 25~MeV bins and low statistics
\cite{PRD79p092001} do not allow for any definite conclusions.

Nevertheless, $R_{b}$ data of the BABAR Collaboration \cite{PRL102p012001}
can be compared to the HORSE predictions for bottomonium
(see Fig.~1 in Ref.~\cite{ARXIV09100967}).
However, in Ref.~\cite{CNPC40p100001}
the non-resonant $B\bar{B}$ threshold enhancement
is classified as the $\Upsilon (4S)$ resonance,
whereas HORSE predicts the central mass of that resonance
to be some 150~MeV heavier.
Threshold enhancements can easily be observed
for $B\bar{B}$, $B^{\ast}\bar{B}$, and $B^{\ast}\bar{B}^{\ast}$
(see Fig.~3 in Ref.~\cite{ARXIV09100967}).
Moreover, non-resonant threshold enhancements for production amplitudes
were predicted in Ref.~\cite{AP323p1215}.
In Fig.~6 in Ref.~\cite{ARXIV09100967}
one observes how the $\Upsilon (4S)$ resonance
interferes with the $B_{s}\bar{B}_{s}$ threshold enhancement.
In Figs.~6 and 7 in Ref.~\cite{ARXIV10094097}
we extracted the $\Upsilon\left(2^{\; 3\!}D_{1}\right)$ state
from data published by the BABAR Collaboration \cite{PRD78p112002}.

Finally, the discovery of a very light hadronic particle, the $E(38)$,
with a mass of about 38~MeV, was not completely unexpected.
A flavour-independent HORSE parameter,
the average $q\bar{q}$/meson-meson interaction distance,
was observed to be related to such a small mass.
But it was not before 25 years later that we became aware
of an interference effect that might be associated with
the existence of a 38-MeV quantum \cite{ARXIV10095191}.
Further evidence \cite{ARXIV11021863}
resulted from leptonic bottomonium decays
published by the BABAR Collaboration \cite{PRD78p112002}.
High-statistics data (see Fig.~8 in Ref.~\cite{ARXIV12021739})
published by the COMPASS Collaboration \cite{ARXIV11086191}
exhibits a very clear diphoton signal,
but the collaboration changed {\it a posteriori}
\/its electronically published article,
claiming that the enhancement is an artifact
of their experimental setup.
However, that claim is not substantiated
by the their own Monte-Carlo simulation
(see Fig.~8 in Ref.~\cite{ARXIV12043287}),
which moreover refers to data with much lower statistics,
published in Ref.~\cite{ARXIV11090272}.

We wish to thank the organizers for the very pleasant workshop.
The present work received partial financial support from FCT Portugal,
with reference no.\ UID/FIS/04564/2016.

\newcommand{\pubprt}[4]{#1 {\bf #2}, #3 (#4)}
\newcommand{\ertbid}[4]{[Erratum-ibid.~#1 {\bf #2}, #3 (#4)]}
\def\AIPCP{{\it AIP Conf.\ Proc.}}
\def\AP{{\it Ann.\ Phys.}}
\def\APPS{{\it Acta Phys.\ Polon.\ Supp.}}
\def\CNPC{{\it Chin.\ Phys.\ C}}
\def\EPJA{{\it Eur.\ Phys.\ J.\ A}}
\def\IJTPGTNO{{\it Int.\ J.\ Theor.\ Phys.\ Group Theor.\ Nonlin.\ Opt.}}
\def\LNP{{\it Lect.\ Notes Phys.}}
\def\NPB{{\it Nucl.\ Phys.\ B}}
\def\PLB{{\it Phys.\ Lett.\ B}}
\def\PRD{{\it Phys.\ Rev.\ D}}
\def\PRL{{\it Phys.\ Rev.\ Lett.}}
\def\PTP{{\it Prog.\ Theor.\ Phys.}}
\def\ZPC{{\it Z.\ Phys.\ C}}

\end{document}